Superconductivity and Disorder in High-$T_c$ Materials:

Crystalline Copper Oxides and Liquid Metallic Hydrogen

W. J. Nellis

Department of Physics

Harvard University

Cambridge, MA 02138

Abstract

Our goals are (i) to assist a non-practitioner toward understanding the likely resolution of the controversy concerning the boson that couples Cooper pairs in superconducting Cu oxides and (ii) to consider if high temperature superconductivity (HTS) in solid oxides might be related to predicted HTS in liquid metallic hydrogen (LMH) near 0 K. LMH is monatomic and both electrons and protons are quantum in nature. I would be astonished if anything contained herein has not been stated previously. High temperatures (T~$T_c$~$\Theta_D$~100 K, where $T_c$ is superconducting critical temperature and $\Theta_D$ is the Debye temperature) induce phonon-phonon scattering in ordered materials, which is very sensitive to impurities and disorder. Experimental results suggest the boson that couples Cooper electron pairs in $Bi_2SrCa_2Cu_2O_8$ (BSSCO), and probably in other high temperature superconductors, is incoherent ion vibration caused by disorder on an atomic scale that is difficult to observe experimentally in most cases. Disorder on a nm scale is observed in BSCCO from ~100 K down to 4 K. Phonons are not well-defined in BSSCO and its superconducting coherence length is comparable to the dimension of its unit cell (~1 nm), as is the case for other oxides. $Sr_2CuO_{3+d}$ ($T_c$=95 K) has a crystal structure that enables experimental demonstration that the nature of disorder of apical oxygen atoms within the unit cell has a strong affect on $T_c$. The basic conclusion is that disorder on an atomic scale within the unit cell and the temperature dependence of phonons at high temperatures (T~ $\Theta_D$) probably must be taken into account in order to understand HTS. Incoherent ion vibrations are analogous to strongly scattered conduction electrons in disordered systems. The latter cause minimum metallic conductivity; the former probably couple Cooper pairs in HTS.

There are several reasons to believe that LMH will be a high temperature superconductor and no known reason why LMH cannot be one.

I. Introduction

My purpose in writing this is to help a non-practitioner toward understanding the likely resolution of a controversy that has been going on for twenty years. Namely, what is the boson that couples Cooper pairs in high-temperature superconductivity (HTS)? The attempt to answer this question has been the focus of extensive research ever since these oxides were discovered [1]. The considerations herein were also motivated by the question as to whether HTS in solid Cu oxides is related to predicted HTS in liquid metallic hydrogen (LMH) at 100 GPa pressures near T=0 K [2]. LMH is monatomic and both electrons and protons are quantum in nature. Its liquid ground state is predicted to be caused by quantum effects. Because ions are disordered in a fluid, the question is equivalent to asking if ion disorder in LMH is related to HTS in solid copper oxides. I would be astonished if anything contained herein has not been stated previously. If anyone else finds the following sufficiently interesting to read, so much the better.

II. Oxides

Sufficient experimental data have now been accumulated to develop a picture of the nature of superconductivity in BSCCO. Atoms in the BSCCO crystal structure are probably disordered sufficiently that phonon frequencies are not well defined, as observed by neutron diffraction below 100 K [3] and as indicated by measurements of thermal conductivity [4]. Disorder in the BSCCO lattice on a nm spatial scale has recently been observed in tunneling experiments at 4 K [5]. Given the essential absence of coherent ion vibrations (phonons) in BSSCO, the boson mediating the attraction between a Cooper electron pair in BSCCO is probably incoherent ion vibration, as suggested by Lee et al [5].

Ion-channeling experiments have suggested ion disorder in YBa$_2$Cu$_3$O$_{7-d}$ (YBCO) [6], as well. Ion disorder in the complex, multi-component, crystal structures of Cu oxides might be a crucial feature for the nature of their superconductivity, including their short coherence lengths [7], and might be a path to obtaining higher T$_c$s.

Our starting point is to first look at low-temperature superconductivity (LTS). Important points to consider are the relative coherence lengths of LTS and HTS and what these imply about the relative length scales of the bosons coupling Cooper pairs. Sn is a representative elemental low temperature superconductor. The crystal structure of white tin is tetragonal with lattice parameters a=b=0.583 nm and c=0.318 nm. The characteristic dimension of the unit cell of Sn is $\ell_U$~0.5 nm. The coherence length $\ell_c$ between electrons of a Cooper pair in Sn is ~200 nm with T$_c$=3.7 K [7]. At 3.7 K only non-interacting, long-wavelength phonons are excited. In Sn a phonon couples a Cooper pair over a distance of ~400 lattice spacings. In order to cause coherence over such a long distance, it is probable that only one phonon wave train is involved.

In high-T$_c$ oxides this situation is very different. A representative coherence length is $\ell_c$ ~1 nm in the a-b plane of BSCCO with T$_c$=95 K [7]. This result suggests that the spatial extent of the boson coupling Cooper pairs in BSCCO also has a spatial extent of ~1 nm. If this boson were coherent over a larger spatial scale, $\ell_c$ would be expected to larger as well.

BSSCO is effectively tetragonal with a=b=0.381 nm and c=3.06 nm. The characteristic dimension of the unit cell of BSSCO is $\ell_U$~1 nm~ $\ell_c$. A natural question suggested by this fact is whether $\ell_c$ is caused by $\ell_U$ or whether this is just a coincidence.

The nature of phonons at ~100 K in BSSCO, discussed below, is much different than those of elemental Sn at 4 K. The decrease in $\xi_c$ by a factor of ~200 and the increase in $T_c$ by a factor of ~20 in BSCCO relative to Sn implies that the boson mediating Cooper pairs is not the same in BSSCO and in Sn. At the same time, both high temperature superconductors (e.g. [5]) and low temperature ones have isotope effects, which implies that the difference in the coupling boson is the nature of ion vibration.

In order for a phonon (coherent ion vibration) to be a meaningful concept, the phonon wavelength must be large compared to the dimension of the unit cell, $\lambda_U$~1 nm for BSSCO. A crude estimate of the shortest phonon wavelength $\lambda_D$ can be obtained using the Debye model. The Debye temperature $\Theta_D$ of BSCCO is 250 K [7]. At an energy corresponding to the Debye temperature, $\Theta_D = hV_{ss}/k_B \lambda_D$, where $h$ is Planck's constant, $V_{ss}$ is the speed of sound, and $k_B$ is Boltzmann's constant. The speed of sound of BSCCO, approximated by the speed of sound of YBCO in the a-b plane, is ~7.0 km/s [8]. Thus, $\lambda_D$ is 1.4 nm, comparable to the dimension of the unit cell $\lambda_U$ (~1 nm). When $\lambda_D \sim \lambda_U$, the concept of a phonon is questionable, though clearly the Debye model is crude for this purpose. However, when the effect of defects, including chemical disorder of elemental atoms within the unit cell of the complex, multi-component, 15-atom basis of the crystal structure of $Bi_2SrCa_2Cu_2O_8$, is considered as well, the concept of a phonon appears to be meaningless, an idea suggested previously [4]. Since the ions do vibrate, these considerations suggest that ion vibrations are incoherent in BSCCO.

Neutron resonances caused by phonons in BSCCO are broadened in energy and their line widths are essentially constant below $T_c$ [3]. Those authors state "Because of the large number of closely spaced phonon modes, it is almost impossible to isolate individual phonon modes." One interpretation of this broadening is that elements in the unit cell are randomly disordered with respect to their ideal positions in the unit cell, independent of temperature below $T_c$. Neutron scattering is a bulk effect.

The existence of disorder in BSSCO motivated measurements of thermal conductivity as a function of temperature below room temperature [4]. Thermal conductivity is sensitive to scattering mechanisms. Two important mechanisms at high temperatures are the scattering of phonons by other phonons and by impurities (including disorder). These can affect substantially the magnitude of maximum thermal conductivity and the temperature of that maximum.

In well-ordered pure materials phonon-phonon scattering is substantial above $\sim \Theta_D/10$, provided Umklapp scattering processes dominate thermal resistivity [9]. In this case, phonon-phonon scattering is responsible for the maximum and subsequent decrease with temperature of thermal conductivity above $\sim \Theta_D/10$. Because phonon-phonon scattering is sensitive to impurities, thermal conductivity is reduced substantially by this scattering mechanism as well. For example, in the case of high-purity Cu, thermal conductivity has a maximum of 130 watts/cm-K at 15 K, a more impure Cu has a maximum thermal conductivity of 8 watts/cm-K at 35 K, and thermal conductivity of a very impure Cu increases monatonically up to 2 watts/cm-K at 200 K [10].

When $T_c \sim 0.4\, \Theta_D \sim 100$ K, phonon-phonon scattering in BSCCO is expected to be significant and very sensitive to impurities, such as chemical disorder. For heat flow primarily within the basal (a-b) plane, the thermal conductivity of BSCCO has a maximum of 0.06 watts/cm-K at 70 K [4]! BSCCO is a very poor thermal conductor, as expected from the neutron data. In fact, [4] has a section entitled "Do phonons exist?" Thermal conductivity is a bulk property. In the limit of strong scattering, the phonon mean-free path $\lambda_{pmfp}$ is expected to be of the order of the dimension of a unit cell $U \sim \lambda_{pmfp}$. In this spatial limit, coherent lattice vibrations do not exist. Rather, ion vibrations incoherent.

Tunneling experiments on BSCCO [5] show that bosons in the electron-boson interaction have energies that are heterogeneous on a ~2 nm spatial scale. No spatially periodic structure was observed in the unprocessed data, which was interpreted in terms of the electron-boson interaction being "spatially scrambled." An oxygen isotope effect is observed in those experiments. Those authors conclude that "uncorrelated lattice vibrations are the probable coupling mechanism." While tunneling measurements look at surfaces, the conclusion from these tunneling experiments is essentially the same one derived from measurements of bulk properties.

Ion-channeling experiments in superconducting YBCO crystals show distortions in the temperature range 300-30 K [6]. Those results were interpreted in terms of uncorrelated static and dynamic distortions associated with Cu-O rows. Those distortions might be caused by chemical disordering within the unit cell and/or imperfect stoichiometry.

Since the unit cell of BSCCO is disordered, the dimension of the unit cell is probably a fundamental limit on the coherence length $\xi_c$. The state of disorder does not improve by considering more unit cells and, thus there is no reason to believe that $\xi_c$ should be larger. A similar statement can probably be made about other Cu oxides, such as YBCO.

$Sr_2CuO_{3+d}$ ($T_c$=95 K) has a structure that facilitates observation of the sensitivity of $T_c$ to disorder of apical oxygen atoms within the unit cell. $Sr_2CuO_{3+d}$ is synthesized by reacting constituent powders at a pressure of 6 GPa at 1100 C for 1 hour and quenching the high pressure phase to ambient. The resulting samples are subjected to various low-temperature anneals that change the nature of the oxygen ordering. The symmetry of the ordering of apical oxygen has a strong effect on measured $T_c$ [11].

The above considerations suggest that both detailed chemical ordering within the unit cell and lattice vibrations at finite temperatures need to be taken into account to understand high-$T_c$ oxides. The fact that $T_c$ is sensitive to crystalline perfection raises the question as to what the intrinsic $T_c$ would be of perfectly-ordered high temperature superconductors. It is possible that their intrinsic $T_c$s are substantially higher than $T_c$s observed to date. It is also possible that without disorder high temperature superconductors do not superconduct. A key question, then, is what state of disorder maximizes $T_c$ in Cu oxide compounds?

Strongly scattered, disorder-induced, incoherent ion vibrations are analogous to strongly scattered conduction electrons in disordered systems, such as fluid metallic hydrogen at finite temperatures. The latter are the cause of minimum metallic conductivity. The former are probably the boson that couples Cooper pairs in HTS.

III. Liquid Metallic Hydrogen

Theoretically calculated $T_c$s (~100 K) of predicted LMH are comparable to those calculated for solid metallic hydrogen [12]. Thus, LMH might be a high temperature superconductor at 100 GPa pressures. Recent theoretical considerations have come to the same conclusion [13].

Degenerate fluid H forms by a continuous transition from a semiconductor to a metal with minimum metallic conductivity at 140 GPa, 9-fold compressed initial liquid density, and ~2600 K [14,15]. At this temperature only the electrons are degenerate and quantum in nature; the protons are classical. This semiconductor-metal transition occurs systematically via a Mott transition in H, N, O, Rb, and Cs [16,17]. Because the temperatures of these fluids are a few 1000 K, these fluids do not superconduct. LMH (quantum protons at low temperatures) is yet to be observed experimentally. Thus, it remains to be seen if fluid metallic hydrogen can be quenched into a dense glass at ambient pressure and if the glass superconducts, as implied by the results in [12].

Atomic disorder in fluid metallic hydrogen, nitrogen, and oxygen causes strong scattering of conduction electrons and minimum metallic conductivities (~500 ohm-cm) at finite temperatures. The normal-state resistivities of Cu oxides are comparable. Quantum disorder in LMH near 100 GPa and 0 K is also expected to produce strong electron scattering and its normal-state resistivity is expected to be a few 100 ohm-cm, as well.

Many disordered systems are known to superconduct. For example, amorphous Bi and Ga become superconducting at $T_c$=~6.0 K and ~7.5 K, respectively [18]. Crystalline Bi has not been observed to superconduct and crystalline Ga superconducts at 1.1 K. The crystalline intermetallic compound $LuRh_4B_4$ superconducts at $T_c$=7 K. Disorder induced by radiation damage by alpha particles decreases its normal-state electrical conductivity to a minimum metallic conductivity of 2500/(ohm-cm) (400 ohm-cm) and decrease $T_c$ to ~1 K [19].

There are several reasons to believe that LMH will be a high temperature superconductor and no known reason why LMH cannot be one.